\documentstyle[epsfig]{l-aa}

\begin{document}

\thesaurus{06(08.01.1; 08.05.3; 08.12.1; 08.16.2)}

\title{Beryllium abundances in parent stars of extrasolar planets: 16 Cyg A \&
B and $\rho^1$ Cnc\thanks{Based on observations made with the Nordic Optical
and William Herschel Telescopes, which are operated on the island of La Palma
by the NOT Scientific Association and the Isaac Newton Group, respectively, in
the Spanish Observatorio del Roque de los Muchachos of the Instituto de
Astrof\'\i sica de Canarias.}}

\author{R. J. ~Garc\'\i a L\'opez\inst{1} \and  M. R. ~P\'erez de Taoro\inst{2}}

\offprints{R. J. ~Garc\'\i a L\'opez (rgl@ll.iac.es)}

\institute{Instituto de Astrof\'\i sica de Canarias, E-38200 La Laguna, 
Tenerife, Spain
\and Museo de la Ciencia y el Cosmos de Tenerife, E-38200 La Laguna, 
Tenerife, Spain}

\date{Received date; accepted date}
\maketitle
\markboth{Garc\'\i a L\'opez \& P\'erez de Taoro: Beryllium in 16 Cyg A \& B
and $\rho^1$ Cnc}{}

\begin{abstract}

The $^9$Be\,{\sc ii} $\lambda$ 3131 \AA\ doublet has been observed in the
solar-type stars 16 Cyg A \& B and in the late G-type star $\rho^1$ Cnc, to
derive their beryllium abundances. 16 Cyg A \& B show similar (solar) beryllium
abundances while 16 Cyg B, which has been proposed to have a planetary
companion of $\sim 2$ $M_{\rm Jup}$, is known to be depleted in lithium by a
factor larger than 6 with respect to 16 Cyg A. Differences in their rotational
histories which could induce different rates of internal mixing of material,
and the ingestion of a similar planet by 16 Cyg A are discussed as potential
explanations. The existence of two other solar-type stars which are candidates 
to harbour planetary-mass companions and which show lithium and beryllium
abundances close to those of 16 Cyg A, requires a more detailed inspection of
the peculiarities of the 16 Cyg system.

For $\rho^1$ Cnc, which is the coolest known object candidate to harbour a
planetary-mass companion ($M > 0.85$ $M_{\rm Jup}$), we establish a precise
upper limit for its beryllium abundance, showing a strong Be depletion which
constrains the available mixing mechanisms. Observations of similar stars
without companions are required to asses the potential effects of the planetary
companion on the observed depletion. It has been recently claimed that $\rho^1$
Cnc appears to be a subgiant. If this were the case, the observed strong
Li and Be depletions could be explained by a dilution process taking place
during its post-main sequence evolution.

\keywords{stars: abundances - stars: evolution - stars: late-type - planetary
systems}
\end{abstract}

\section{Introduction} 
\label{sec1} 

In very recent years, several stars have been proposed to have planetary
companions on the basis of measured precise radial velocity variations. This
field of research is experiencing rapid development, and updated reviews of the
present situation can be found in the proceedings of the workshop on {\it Brown
Dwarfs and Extrasolar Planets} edited by Rebolo et al. (1998) and in The
Extrasolar Planets Encyclopaedia\footnote{http://wwwusr.obspm.fr/planets/} by
J. Schneider. 

Once a solar-type star has been suggested to harbour a planetary-mass
companion, it is interesting to investigate any similarities with the Sun, as
well as to find possible differences with respect to other single stars.
Chemical abundances are among the most important parameters to be compared and,
in particular, precise abundances of light elements such as lithium and
beryllium (easy to destroy by $(p,\alpha)$ nuclear reactions when the
temperature reaches $\sim 2.5\times 10^6$ and $\sim 3.5\times 10^6$ K,
respectively) combined with the abundances of other elements which are not so
readily destroyed in stellar interiors, should help to understand how the
presence of planets may affect the chemical composition of their parent stars.
Gonzalez (1997, 1998) has derived the overall metallicities as well as
abundances of different elements (including lithium) for a wide sample of
proposed parent stars, finding that four of the known systems show  a
metallicity significantly higher than the solar value. 

A peculiar system such as 16 Cyg A \& B, formed by twin solar-type stars of
which only one has an orbiting planet (Cochran et al. 1997), is an especially
suitable candidate to perform a detailed abundance study. Gonzalez (1998) found
that both stars have a similar metallicity with a value slightly larger than
solar, and confirmed independently a previous result of King et al. (1997a)
that 16 Cyg B (the star with a suspected planet) is strongly depleted in
lithium with respect to 16 Cyg A. The knowledge of their beryllium abundances
is of potential value in quantifying the possible influence of a planetary
companion on the mixing mechanisms operating in the stellar interior.

$\rho ^1$ Cnc is a star with spectral type G8V, and is the coolest known object
which is a candidate to have a planetary companion. Following Gonzalez (1998),
this star falls into the group having roughly Jupiter-mass companions with
small circular orbits and very metal-rich parent stars. Dominik et al. (1998)
have shown recently that the planetary system of $\rho ^1$ Cnc also hosts a
Vega-like disk of dust, evidenced by an infrared excess at 60 $\mu$m. The star
is very depleted in lithium and its beryllium abundance could be compared with
existing upper limits measured in younger stars with similar effective
temperatures (Garc\'\i a L\'opez et al. 1995a). 

In this paper we derive the beryllium abundances of the 16 Cyg system and of
$\rho ^1$ Cnc by comparing observations with spectral syntheses of the
$^9$Be\,{\sc ii} $\lambda$ 3131 \AA\ doublet. We use those, together with their
published lithium values, as well as with available abundances for other stars
(with and without suggested planetary companions), and discuss briefly possible
effects of planets on processes taking place in their structure and evolution.

\section{Observations and data reduction}
\label{sec2}

The observations were carried out in several runs conducted at the 2.6 m Nordic
Optical (NOT) and 4.2 m William Herschel (WHT) telescopes of the Observatorio
del Roque de los Muchachos (La Palma), using the IACUB (McKeith et al. 1993)
and Utrecht Echelle (UES) spectrographs, respectively. Table 1 lists the stars
observed, dates, and telescopes. Most of the observations were performed at the
NOT and only two exposures of 1200 s each were devoted to 16 Cyg B at the WHT.
$\rho ^1$ Cnc and 16 Cyg A were observed with IACUB using a slit width of $\sim
0.7$ arc sec which provided spectra with a resolution $R=\lambda
/\Delta\lambda\sim 40000$ recorded using a $1024\times 1024$ (19 $\mu$m)
Thomson blue-coated CCD, while the slit width used to observe 16 Cyg B at the
NOT was $\sim 0.8$ arc sec and the resolution $R\sim 33000$. IACUB is an
un-crossed echelle spectrograph and the order corresponding to the Be\,{\sc ii}
doublet was isolated using an interference filter centered at $\lambda$ 3131
\AA, with a FWHM of 45 \AA, and a maximum transmission of 64 \%. The UES
observations of 16 Cyg B were performed using the E79 grating, a slit width of
$\sim 1$ arc sec, $R\sim 50000$, and recorded with a $2048\times 2048$ (24
$\mu$m) SITe CCD. Eleven spectral orders were available on the detector, the
bluest one with useful signal containing the Be\,{\sc ii} doublet.

\begin{table*}
\caption[]{Stars observed}
\begin{flushleft}
\begin{tabular}{lccccccc}
\hline
\noalign{\vspace {0.1cm}}
 Star & Name & $V$ & $B-V$ & Telescope & Date & Exp. time (s) & S/N\\
\noalign{\vspace {0.1cm}}
\hline
\noalign{\vspace {0.1cm}}
HR 3522 & $\rho^1$ Cnc & 5.95 & 0.87 & NOT & 19/11/1996 & 3600 & 35\\
HR 7503 & 16 Cyg A     & 5.96 & 0.64 & NOT & 19/07/1997 & 1800 & 35\\
HR 7504 & 16 Cyg B     & 6.20 & 0.66 & NOT & 22/07/1997 & 1800 & 30\\
        &              &      &      & WHT & 13/11/1997 & 2400 & 15\\
\noalign{\vspace {0.1cm}}
\hline
\end{tabular}
\end{flushleft}
\label{table1}
\end{table*}

Data reductions were performed by standard procedures using routines included
in the IRAF\footnote{IRAF is distributed by National Optical Astronomy
Observatories, which is operated by the Association of Universities for
Research in Astronomy, Inc., under contract with the National Science
Foundation, USA.} suite of programs. Final signal-to-noise ratios (S/N)
achieved in the pseudo-continuum surrounding the Be\,{\sc ii} doublet are
listed in Table 1. Technical problems with the UES and the low elevation of the
object  during the observations (airmass $\sim 2-3$) prevented the achievement
of a better spectrum of 16 Cyg B with the WHT. Due to the absence of ThAr lines
available for these observations, wavelength calibrations were carried out
using photospheric lines present in the stellar spectra whose values were taken
from the Moore et al. (1966) solar atlas. Second-order polynomial fits where
applied, using 11--18 lines, to give calibrations with rms scatter between
0.004 and 0.011 \AA.  Dispersions of 0.017 and 0.035 \AA\ pixel$^{-1}$ were
obtained for $\rho ^1$ Cnc and 16 Cyg A \& B, respectively, with IACUB (with a
binning of two pixels in the spectral direction for the latter cases), while a
dispersion of 0.033 \AA\ pixel$^{-1}$ corresponded to the UES spectrum of 16
Cyg B.

\section{Spectral synthesis}
\label{sec3}

Stellar parameters were taken from the detailed abundance analysis carried out
by Gonzalez (1998). Adopted effective temperatures ($T_{\rm eff}$), surface
gravities ($\log g$), and metallicities ([Fe/H]) are listed in Table 2. Typical
errors associated with these quantities are $\pm 100$ K for $T_{\rm eff}$ and
$\pm 0.1$ dex for both $\log g$ and [Fe/H]. A solar oxygen abundance log
N(O)$=8.93$ (Anders \& Grevesse 1989; where log N(X)$=$log (X/H)$+$12) was
taken, in all cases. Small changes of this abundance (like the updated solar
value log N(O)$=8.87$ by Grevesse \& Noels 1996), which affect the OH lines
located in the Be\,{\sc ii} region, do not alter significantly the comparison
between observed and synthetic spectra.  

The beryllium abundance analysis was performed by spectral synthesis fitting to
the $^9$Be\,{\sc ii} $\lambda$ 3131.065 \AA\ line, which is weaker but more
isolated than its companion at $\lambda$ 3130.421 \AA. Synthetic spectra were
computed using the code WITA2, a UNIX-based version of code ABEL6 (Pavlenko
1991), which computes LTE atomic and molecular line profiles, and the model
photospheres were interpolated for the adopted stellar parameters within a grid
of ATLAS9 models (computed without overshooting) provided by Kurucz (1992).
Details of the line list employed (which is adjusted to reproduce the solar
spectrum and uses accurate oscillator strengths for the Be\,{\sc ii} lines),
the absence of significant NLTE effects on the derived abundance ($<0.1$ dex),
and the low sensitivity of the observed feature to changes in the beryllium
abundance for low $T_{\rm eff}$ values were shown in Garc\'\i a L\'opez et al.
(1995a,b).

\subsection{16 Cyg A and B}
\label{sec3.1}

The values of the parameters adopted for 16 Cyg A \& B are very close to the 
solar values. Figure 1 shows the comparison between observed and synthetic
spectra for the Sun and 16 Cyg A \& B. The solar spectrum, with a resolution
$R\sim 50000$, was obtained by observing the Moon with the combination
WHT$+$UES in a previous campaign during April 1995. Synthetic spectra have been
convolved with gaussians with the appropriate FWHMs to reproduce the different
instrumental profiles. A beryllium abundance log N(Be) $=1.15$ (Chmielewski et
al. 1975; Anders \& Grevesse 1989) reproduces very well the observed solar
$^9$Be\,{\sc ii} $\lambda$ 3131.065 \AA\ line, while this is not the case for
other surrounding lines (including the $^9$Be\,{\sc ii} $\lambda$ 3130.420 \AA\
line), and provides a fiducial comparison for the quality of the best fit which
can be achieved for solar-type stars. Garc\'\i a L\'opez et al. (1995b)
illustrate the high sensitivity of solar synthetic spectra to changes in the
beryllium abundance. The best fit to the observed weak Be\,{\sc ii} line for 16
Cyg A indicates an abundance of log N(Be) $=1.10$, i.e. a beryllium abundance
not significantly different from solar. Changes of $\pm 250$ K in $T_{\rm
eff}$, $\pm 0.3$ dex in $\log g$, and $\pm 0.2$ dex in [Fe/H] imply variations
of $\pm 0.05$, $\pm 0.2$, and $\pm 0.05$ dex, respectively, in the solar Be
abundance (Garc\'\i a L\'opez et al. 1995b). The corresponding maximum
uncertainties for the parameters adopted for 16 Cyg A are $\pm 0.05$, $\pm
0.1$, and $\pm 0.05$ dex, respectively. Including an additional 0.05 dex
abundance error induced by the uncertainty in locating the pseudo-continuum
around the Be\,{\sc ii} doublet, and another 0.1 dex induced by the uncertainty
associated with the available S/N, the final error resulting from combining
these errors in quadrature amounts $\pm 0.17$ dex. 

\begin{table*}
\caption[]{Stellar parameters and lithium \& beryllium abundances}
\begin{flushleft}
\begin{tabular}{cccccc}
\hline
\noalign{\vspace {0.1cm}}
 Star & $T_{\rm eff}$ (K) & $\log g$ & [Fe/H] & log N(Li) & log N(Be) \\
\noalign{\vspace {0.1cm}}
\hline
\noalign{\vspace {0.1cm}}
$\rho ^1$ Cnc & 5150$^a$ & 4.15$^a$ & \phantom{$-$}0.29$^a$ &            
$<-$0.04$^a$ & $< 0.55^b$ \\
16 Cyg A      & 5750$^a$ & 4.20$^a$ & \phantom{$-$}0.11$^a$ &  
\phantom{$<-$}1.24$^a$ & $1.10\pm 0.17^b$ \\
16 Cyg B      & 5700$^a$ & 4.35$^a$ & \phantom{$-$}0.06$^a$ &
$<$\phantom{$-$}0.46$^a$ & $1.30\pm 0.17^b$ \\
70 Vir        & 5538$^c$ & 4.02$^c$ &           $-$0.04$^c$ &  
\phantom{$<-$}1.79$^c$ & $0.86\pm 0.22^c$ \\
HD 114762     & 5865$^c$ & 4.31$^c$ &           $-$0.66$^c$ &  
\phantom{$<-$}1.94$^c$ & $0.95\pm 0.33^c$ \\
$\upsilon$ And& 6050$^d$ & 4.0~$^d$ & \phantom{$-$}0.06$^d$ &
\phantom{$<-$}2.15$^d$ & \phantom{$<$} 0.90$^d$ \\
$\tau$ Boo    & 6390$^d$ & 3.8~$^d$ & \phantom{$-$}0.30$^d$ &
$<$\phantom{$-$} 0.6\phantom{0}$^d$ & $<0.05^d$ \\
\noalign{\vspace {0.1cm}}
\hline
\end{tabular}
\end{flushleft}
Data taken from Gonzalez (1998; $a$), this work ($b$), Stephens et al.
(1997; $c$), and Boesgaard \& Lavery (1986; d).
\label{table2}
\end{table*}

\begin{figure*}
\mbox{\epsfig{file=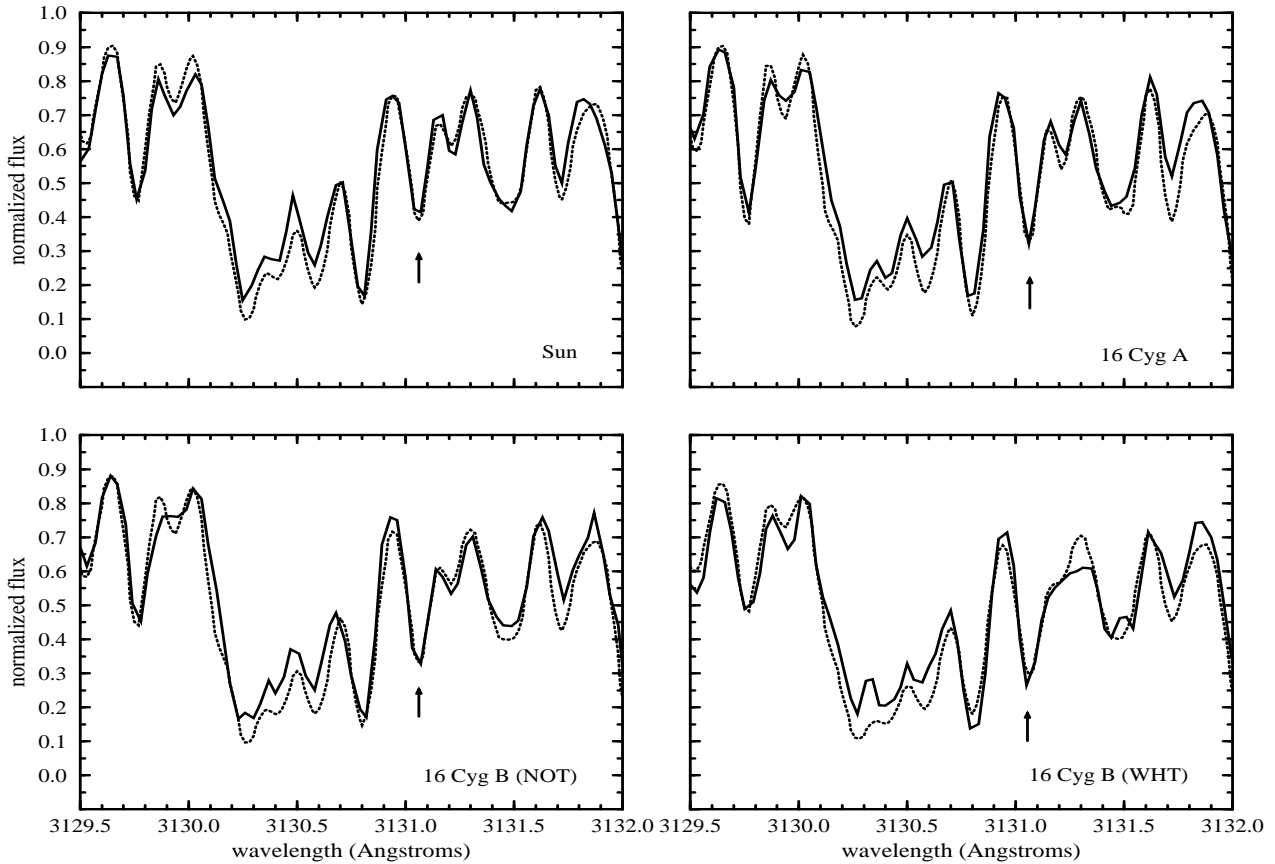,height=12.cm,width=17.cm,angle=270}}
\caption[]{Best fit of synthetic spectra (dotted line) to the observations
(solid line) of the Sun and 16 Cyg A \& B. Two independent spectra of 16 Cyg B
obtained at the NOT and WHT, respectively, are shown in the lower panels.
Beryllium abundances have been derived from the fitting to the (indicated)
$^9$Be\,{\sc ii} $\lambda$ 3131.065 \AA\ line.}
\label{fig1}
\end{figure*}

The beryllium abundance derived from a similar analysis applied to the spectrum
of 16 Cyg B observed with NOT$+$IACUB is log N(Be) $=1.30\pm 0.17$, slightly
larger than that derived for 16 Cyg A but compatible with it and the solar
value within the error bars. This shows that while there is a difference of a
factor 6 (at least) in the lithium abundances of both stars (listed in Table 2,
and taken from Gonzalez 1998), there is no indication of beryllium depletion
among them. A service observation of 16 Cyg A \& B was requested at the WHT
aimed at improving the S/N obtained with the NOT and better constraining the
slight difference in Be abundance between the two stars. However, technical
problems with the UES and the restriction imposed at the time of the
observation by the coordinates of the objects allowed us to obtain only one
final low-S/N spectrum for 16 Cyg B. The UES spectrum shown in Fig. 1 has been
smoothed slightly, and the synthetic spectrum overplotted was computed with an
abundance of log N(Be) $=1.6$.

\subsection{$\rho ^1$ Cnc}
\label{sec3.2}

The low effective temperature of this star makes its beryllium abundance
analysis more difficult and uncertain. Garc\'\i a L\'opez et al. (1995a)
studied the sensitivity of the observed Be\,{\sc ii} $\lambda$ 3131.065 \AA\
feature to the beryllium abundance in late-type stars belonging to the Hyades
open cluster and the Ursa Major Group (UMaG), and found that this sensitivity
decreases with decreasing $T_{\rm eff}$. A line of another element which is
blended with the Be line becomes very important for cool stars, clearly
dominating the feature when the temperature drops below $\sim 5000$ K. They
tentatively identified the perturbing line as Mn\,{\sc i} $\lambda$ 3131.037
\AA, and Primas et al. (1997) and King et al. (1997b) also found evidence of
such a blend in their analyses of $\alpha$ Cen A \& B, but suggesting 
different blending features. As a result of this limitation, Garc\'\i a L\'opez
et al. derived reliable beryllium abundances only for three Hyades stars with
$5700\ge T_{\rm eff}\ge 5200$ K, and established upper limits for the cooler
stars in their sample. On the other hand, Garc\'\i a L\'opez (1996)
investigated the potential use of $^9$Be\,{\sc i} lines to overcome the
uncertainties associated with the Be\,{\sc ii} line at low effective
temperatures, finding that the best candidate ($\lambda$ 2348.609 \AA, not
observable from the ground) would not provide reliable abundances.

Figure 2 shows the comparison between the observed spectrum of $\rho ^1$ Cnc
and several synthetic spectra computed with different Be abundances and the
stellar parameters listed in Table 2. As seen in the upper panel, the fit of
the synthetic spectrum to the observed one is not as good as for solar-like
stars (Fig. 1); a value log N(Be) $= 0.1$ reproduces the observed feature. The
lower panel is a zoom of the region surrounding the Be\,{\sc ii} $\lambda$
3131.065 \AA\ line, and the observed spectrum is represented by photon
statistics error bars to better demonstrate the sensitivity of the observed
feature to changes in beryllium abundance. Five synthetic spectra, computed
without beryllium and with log N(Be) $=0.1$, 0.3, 0.5, and 0.7, respectively,
are also shown. Although the spectrum computed with log N(Be) $=0.1$ reproduces
the observations, the spectrum without beryllium is well included within the
error bars suggesting that an upper limit instead of a measurement is a more
prudent result here. This upper limit could be as high as log N(Be) $=0.5$ given
the S/N in the points defining the line itself. An additional 0.15 dex
uncertainty is induced by the errors in the adopted stellar parameters (mainly
$\log g$), 0.1 dex is associated with the uncertainty in locating the
pseudo-continuum, and, finally, we consider an error of 0.1 dex related to an
uncertainty of $\pm 0.2$ dex in the estimated $\log gf$ value for the Mn\,{\sc
i} line which blends strongly the Be\,{\sc ii} line at this effective
temperature. Combining these uncertainties in quadrature, our conservative
upper limit for the beryllium abundance of $\rho ^1$ Cnc is log N(Be) $<0.55$.

\begin{figure}
\mbox{\epsfig{file=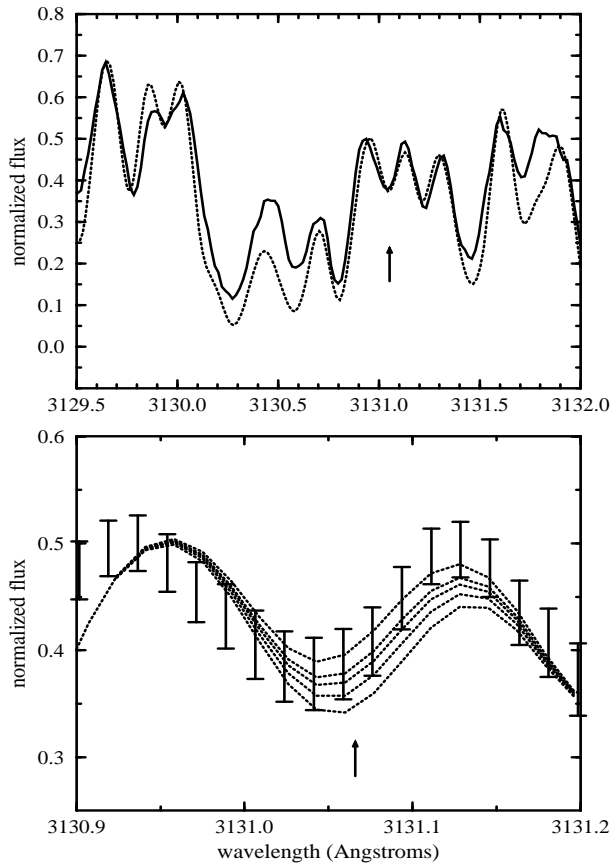,height=12.cm,width=15.8cm,angle=270}}
\caption[]{Upper panel: best fit of a synthetic spectrum (dotted line) to the
observations of $\rho ^1$ Cnc, indicating an abundance log N(Be) $=0.1$.
Lower panel: zoom of the region surrounding the Be\,{\sc ii} $\lambda$ 3131.065
\AA\ line. The observed spectrum is represented by photon statistics error
bars. Five synthetic spectra (dotted lines), computed without beryllium and log
N(Be) $=0.1$, 0.3, 0.5, and 0.7, respectively, show the low sensitivity of the
observed feature to changes in the beryllium abundance of the synthetic
profiles, and indicate that only an upper limit can be derived.}
\label{fig2}
\end{figure}

\section{Discussion}
\label{sec4}

\subsection{16 Cyg A and B}
\label{sec4.1}

The lithium and beryllium abundances of 16 Cyg A are very similar to those of
the Sun and $\alpha$ Cen A, another well known solar analog. Adopting $T_{\rm
eff}=5800$ K, $\log g=4.31$, and [Fe/H]$=+0.24$ for $\alpha$ Cen A (Chmielewski
et al. 1992), King et al. (1997a) found log N(Li) $=1.37\pm 0.06$ and King et
al. (1997b) derived log N(Be) $=1.32\pm 0.15$ for this star. Primas et al.
(1997) obtained log N(Be) $=1.21\pm 0.09$ using the same effective temperature
and slightly different gravity and metallicity ($\log g=4.40$ and
[Fe/H]$=+0.10$). Stephens et al. (1997) derived Li and Be abundances for a
large sample of F- and G-type main-sequence (MS) stars. Two of the stars
studied by Stephens et al. have also been proposed to have very-massive
planetary companions: 70 Vir (HR 5072) and HD 114762, with $M_{\rm planet}>9$
$M_{\rm Jup}$ and $>10$ $M_{\rm Jup}$, respectively, following Gonzalez (1998).
The effective temperatures and Li \& Be abundances assigned by Stephens et al.
to these stars are listed in Table 2, and it can be seen that they do not show
the same dramatic Li depletion as does 16 Cyg B. There are two additional 
parent stars with available Be abundances: $\upsilon$ And and $\tau$ Boo,
studied by Boesgaard \& Lavery (1986) using photographic spectra. These stars
are the hottest parent stars found up to now. $\tau$ Boo, with an effective
temperature of 6390 K assigned by Boesgaard \& Lavery, is located well 
into the so called ``Li gap'' of the F-type stars (Boesgaard \& Tripicco 1986)
and shows significant Li and Be depletions, while $\upsilon$ And (with $T_{\rm
eff}\sim 6050$ K and close to the red edge of the gap) has kept large 
amounts of both elements (see Table 2). 

Most of the stars studied by Stephens et al. in the interval $T_{\rm eff}\sim
5700-5800$ K show lithium abundances in the range log N(Li) $\sim 1.6$ to 2.4
and beryllium abundances between 1.0 and 1.3; two stars show Li abundances at
1.41 and 1.71, and some Be depletion at $0.68\pm 0.26$ and $0.86\pm 0.13$,
respectively; and there is also one star, HD 160693, which appears to be
depleted in both Li ($<1.14$) and Be ($0.42\pm 0.22$). The only two objects in
that sample which show somewhat similar behaviour to 16 Cyg B are HR 483 and HR
1729. These have solar beryllium abundance but are significantly hotter
($T_{\rm eff}\sim 5860$ and 5930 K, respectively) and have very-high upper
limits for lithium (log N(Li) $<1.85$ and $<1.76$), which make them also
compatible with an object like 16 Cyg A. This distribution of abundances can be
seen in Figure 3, where we have plotted Be against Li abundances for stars
cooler than $\sim 6060$ K (to avoid the large scatter observed for the
abundances of the hotter stars -Stephens et al. 1997; Garc\'\i a L\'opez et al.
1998-), including $\alpha$ Cen A \& B as well as the Hyades and UMaG stars
studied by Garc\'\i a L\'opez et al. (1995a). Tables 2 and 3 list the data used
to produce Fig. 3. Note that there is a difference of about three orders of
magnitude in the lithium abundances plotted in the figure, while the
corresponding beryllium range is $\sim 1$ dex. Although the stellar parameters
employed to derive the abundances for the different samples are not homogeneous
and detailed star to star comparisons cannot be made, Fig. 3 shows that there
are not significant differences between stars with and without massive
planetary companions, with the exception of 16 Cyg B.

\begin{figure}
\mbox{\epsfig{file=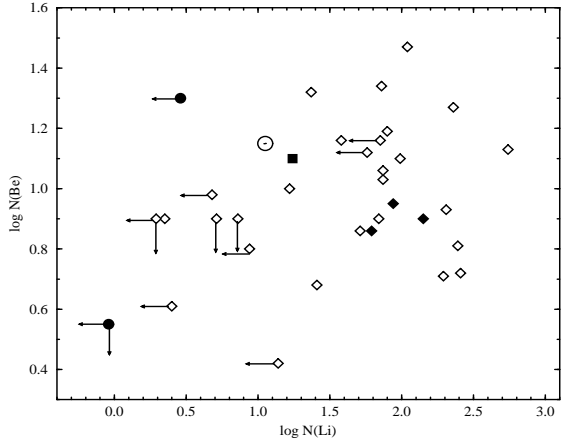,height=7.cm,width=9.cm,angle=270}}
\caption[]{Beryllium against lithium abundances for stars cooler than $\sim
6060$ K. Filled circles: parent stars with Be abundances derived in this work;
filled square: 16 Cyg A; filled diamonds: parent stars with Be abundances taken
from Stephens et al. (1997) and Boesgaard \& Lavery (1986); open diamonds:
single stars with Be abundances taken from the literature  (Garc\'\i a L\'opez
et al. 1995a; King  et al. 1997a,b; Primas et al. 1997;  Stephens et al. 1997);
the solar abundances are marked with the standard symbol.}
\label{fig3}
\end{figure}	

\begin{table}
\caption[]{Stars without massive planetary companions plotted in Figure 3}
\begin{flushleft}
\begin{tabular}{lcccc}
\hline
\noalign{\vspace {0.1cm}}
 Star & $T_{\rm eff}$ (K) & log N(Li) & log N(Be) & Reference\\
\noalign{\vspace {0.1cm}}
\hline
\noalign{\vspace {0.1cm}}
HR 219	       & 5883 &	\phantom{$<$}1.90 & \phantom{$<$}1.19 & 1   \\
HR 483	       & 5862 &	$<$1.85           & \phantom{$<$}1.16 & 1   \\
HD 30649       & 5716 &	\phantom{$<$}1.71 & \phantom{$<$}0.86 & 1   \\
HR 1729	       & 5931 &	$<$1.76           & \phantom{$<$}1.12 & 1   \\
HR 3064	       & 5941 &	\phantom{$<$}2.04 & \phantom{$<$}1.47 & 1   \\
HD 65583       & 5328 &	$<$0.94           & \phantom{$<$}0.80 & 1   \\
HR 4845        & 5830 &	\phantom{$<$}1.87 & \phantom{$<$}1.06 & 1   \\
HR 5914	       & 5801 &	\phantom{$<$}2.36 & \phantom{$<$}1.27 & 1   \\
HR 6060	       & 5809 &	\phantom{$<$}1.58 & \phantom{$<$}1.16 & 1   \\
HD 148816      & 5833 &	\phantom{$<$}1.84 & \phantom{$<$}0.90 & 1   \\
HR 6189	       & 6058 &	\phantom{$<$}2.41 & \phantom{$<$}0.72 & 1   \\
HR 6349	       & 6028 &	\phantom{$<$}2.74 & \phantom{$<$}1.13 & 1   \\
HD 157089      & 5739 &	\phantom{$<$}1.87 & \phantom{$<$}1.03 & 1   \\
HD 160693      & 5701 &	$<$1.14           & \phantom{$<$}0.42 & 1   \\
HR 6775	       & 5955 &	\phantom{$<$}2.31 & \phantom{$<$}0.93 & 1   \\
HD 18499       & 5725 &	\phantom{$<$}1.41 & \phantom{$<$}0.68 & 1   \\
HD 195633      & 5864 &	\phantom{$<$}2.29 & \phantom{$<$}0.71 & 1   \\
HD 208906      & 5940 &	\phantom{$<$}2.39 & \phantom{$<$}0.81 & 1   \\
HR 9088	       & 5377 &	$<$0.68           & \phantom{$<$}0.98 & 1   \\
HR 9107	       & 5574 &	\phantom{$<$}1.86 & \phantom{$<$}1.34 & 1   \\
$\alpha$ Cen A & 5800 &	\phantom{$<$}1.37 & \phantom{$<$}1.32 & 2,3 \\
$\alpha$ Cen B & 5350 &	$<$0.40           & \phantom{$<$}0.61 & 4,5 \\
vB 17	       & 5635 &	\phantom{$<$}1.99 & \phantom{$<$}1.10 & 6   \\
vB 21	       & 5250 &	\phantom{$<$}0.35 & \phantom{$<$}0.90 & 6   \\
vB 26	       & 5465 &	\phantom{$<$}1.22 & \phantom{$<$}1.00 & 6   \\
vB 46	       & 5065 & $<$0.29           & $<$0.90           & 6   \\
HD 41593       & 5140 &	\phantom{$<$}0.86 & $<$0.90           & 6   \\
HD 109011A     & 4760 &	\phantom{$<$}0.71 & $<$0.90           & 6   \\
HD 110463      & 4800 &	\phantom{$<$}0.71 & $<$0.90           & 6   \\
\noalign{\vspace {0.1cm}}
\hline
\end{tabular}
\end{flushleft}
References: (1) Stephens et al. (1997); (2) King et al. (1997a); (3) King et
al. (1997b); (4) Chmielewski et al. (1992); (5) Primas et al. (1997); (6)
Garc\'\i a L\'opez et al. (1995a).
\label{table3}
\end{table}

Gonzalez (1998) estimated an age of $9\pm 2$ Gyr for 16 Cyg A \& B using 
stellar evolution grids. If the difference in lithium abundances between the
two stars (which have kept a similar Be abundance) is caused by a slow mixing
mechanism, which transports the material from the base of the outer convection
zone to the Li burning layer, associated with the angular momentum loss (as
suggested by King et al. 1997a), the stars must have had a different angular
momentum history (i.e. different initial angular momenta or different loss
rates). This would also be the case if the differences in lithium abundance
were related to changes in the stellar structure induced by different rotation
rates during the pre-MS phase (Mart\'\i n \& Claret 1996). Cochran et al.
(1997) suggest that the difference in rotation rate or angular momentum loss
between the two stars could be associated with the existence of a massive
proto-planetary disk around 16 Cyg B while the star was in its pre-MS phase
(giving rise to the present $\sim 2$ $M_{\rm Jup}$ orbiting planet), which
would have served to brake the stellar rotation and induced a more rapid Li
depletion compared to 16 Cyg A, which would have lacked a similar disk and kept
a faster rotation rate. Apart from explaining how a system such as 16 Cyg A \&
B, with only one star surrounded by a massive proto-planetary disk can be
formed, one must also show why the Li depletion is not so severe in 70 Vir and
HD 114762 with more massive planetary companions and greater ages ($8.5\pm 1$
and $14\pm 2$ Gyr, respectively; Gonzalez 1998). A possible explanation could
lie in the large eccentric orbits of the planetary companions to these stars,
with proto-planetary disks which could have then had a smaller effect on the
pre-MS stellar rotation rates. Although the companion to 16 Cyg B also presents
a very-high orbital eccentricity, it is possible that it has been altered from
a more circular one by gravitational perturbation due to 16 Cyg A (Mazeh et al.
1997). In addition, a check has to be made on whether or not the mixing
mechanisms proposed predict correctly the observed abundances. From Figure 11
of Stephens et al. (1997), it seems that the rotationally-induced model of
Deliyannis \& Pinsonneault (1993) predicts significant Li differences and
similar (solar) beryllium abundances for stars with $T_{\rm eff}\sim 5700$ K,
an age of 4 Gyr, and a difference of 20 km s$^{-1}$ in their initial equatorial
velocity. However, it is not clear from the figure what would be the
predictions for stars of considerably greater ages (in the range $\sim 8-14$
Gyr). 

Furthermore, long-period orbiting planets of small mass, like the planets of
the solar system, could also be able to induce some lithium depletion but are
not easy to detect with the currently available observing facilities,
making so very difficult to complete separate a priori the stars with and
without ``Li-depleting'' planets. 

Alternatively, the fact that we do not find stars with lithium and beryllium 
abundances comparable to those of 16 Cyg B (while it is common to find
other stars with orbiting planets behaving like 16 Cyg A) could be telling us
something about the chemical enrichment of stars with planetary companions.
Orbital migration of planets formed at large radii and transported inwards by
tidal interactions with the proto-planetary disk could occur (Lin et al. 1996),
and would result in adding metal-rich material to the parent star. The degree
of chemical enrichment would depend on the fraction of the star over which the
accreted material is distributed, which is linked to the age and evolutionary
stage of the parent star. Laughlin \& Adams (1997) predict that solar-type
stars with maximum disk lifetimes of $\sim 10$ Myr should have virtually no
metallicity enhancement, while more massive stars (early F- to late A-type,
with shallower convection zones) could experience more significant chemical
enrichments. Gonzalez (1998) discusses the possibility that 16 Cyg A could have
had in the past a planetary companion like that of 16 Cyg B, but which was
engulfed by its parent star after the time needed to disturb the companion of
16 Cyg B from its original circular orbit. This could have increased its
observed lithium abundance (which survives in the outer fraction of the stellar
interior where the temperature is lower than $\sim 2.5\times 10^6$ K) without
changing significantly ($<0.1$ dex) the overall stellar metallicity, but what
would be its effect on the beryllium abundance? It will be interesting to
perform detailed computations simulating the accretion of planets of different
sizes onto parent stars with different masses at different evolutionary stages,
to obtain better estimates of the possible enrichment of light elements.
However, if this were the  explanation of the abundances observed for 16 Cyg A
\& B, what would be the peculiarities of this system which cause it to differ
from parent stars with similar effective temperatures, such as 51 Peg, 47 UMa,
HD 114762, the Sun, or even the cooler 70 Vir and hotter $\upsilon$ And, which 
do not show such dramatic lithium depletion? New interesting challenges are 
open for detailed stellar structure and evolution studies.

\subsection{$\rho ^1$ Cnc}
\label{sec4.2}

Although only an upper limit, to the best of our knowledge this value
represents the first evidence of significant beryllium depletion in the 
coolest MS stars measured. King et al. (1997b) and Primas et al. (1997) derived
log N(Be) $<1.17$ and log N(Be) $=0.61\pm 0.28$ for $\alpha$ Cen B adopting,
respectively, $T_{\rm eff}=5325$ and 5350 K, while the upper limit established
by Garc\'\i a L\'opez et al. (1995a) for one Hyades and three UMaG late-type
stars with $4760\ge T_{\rm eff}\ge 5140$ K was log N(Be) $<0.9$, close to the
solar abundance. The effective temperature adopted by Garc\'\i a L\'opez et al.
for the UMaG star HD 41593 is the same as that adopted here for $\rho ^1$ Cnc.
The analysis of HD 41593 shows that a similar low beryllium abundance is
compatible with the observed spectrum, but its low S/N ($\sim 15$) makes it
also compatible with a larger abundance. 

The stellar parameters derived by Gonzalez (1998) for $\rho ^1$ Cnc place the
star in the subgiant region of the HR diagram, confirming previous claims in
this direction. However, the age derived for the star using theoretical
isochrones turns to be $\gg 16$ Gyr, i.e. much greater than the accepted age of
the universe. If $\rho ^1$ Cnc were indeed a subgiant star, the observed Li and
Be depletions could be due to a dilution effect taking place once the star has
evolved off the MS and the convection zone deepens and mixes Li- and Be-rich
material with  Li- and Be-free material from the inner region. Strong Li and Be
depletions have been found by Boesgaard \& Chesley (1976) among several late G-
and early K-type subgiants, which appear to agree with theoretical predictions
for dilution.

A possible alternative explanation suggested by Gonzalez (1998) is that $\rho
^1$ Cnc is an unresolved stellar binary viewed nearly pole-on, and this could
be tested by searching for variations in the line profile shapes. The very-low
lithium abundance of $\rho ^1$ Cnc is compatible with strong depletion
experienced by an old late G-type MS star ($\sim 5$ Gyr, Baliunas et al. 1997
based on Ca\,{\sc ii} H \& K chromospheric activity), as predicted by
extrapolating the 4 Gyr Li isochrones of the  rotating models of Deliyannis \&
Pinsonneault (1993) presented in Fig. 11 of Stephens et al. (1997).
Extrapolating the corresponding Be isochrones in the figure may not give an
adequate prediction of the Be abundance for $\rho ^1$ Cnc, because strong Be
depletion may occur for effective temperatures much cooler than those
considered in the plot. Indeed, older rotating models computed by Pinsonneault
et al. (1990; case A), which include angular momentum loss, predict beryllium
depletions of 0.23 to 0.45 dex for 0.8--0.9 $M_\odot$ stars ($T_{\rm eff}\sim
5000-5400$ K) at 1.7 Gyr (the greatest age considered), with initial angular
momenta in the range $1.6\times 10^{49} - 5\times 10^{50}$ gr cm$^{-2}$
s$^{-1}$. Other angular momentum loss and internal redistribution properties
(cases labeled as B to F) predict similar Be depletions at that age. The strong
Be depletion observed in $\rho ^1$ Cnc would therefore set a very significant
constraint on the theoretical models. Other old and cool stars which are not
suspected to have a companion planet should be observed to distinguish any
possible planetary influence on the Be abundance of $\rho ^1$ Cnc.

\section{Conclusions}
\label{sec5}

Beryllium abundances have been derived for the solar-like stars 16 Cyg A \& B
and the cooler object $\rho ^1$ Cnc, for which there are published values of
their lithium abundances. 16 Cyg B and $\rho ^1$ Cnc are candidates to be
parents of extrasolar planets, and by measuring their Be abundances we aim at
studying the potential dependence on the presence of planetary companions of
detailed processes operating in their structure and evolution.

16 Cyg A \& B show very similar Be abundances, which are compatible with the
solar value, while the lithium abundance of 16 Cyg B is at least a factor 6
smaller than that of 16 Cyg A. Different rates of mixing of material in their
interiors associated with different angular momentum histories, as well as the
hypothetical ingestion of a planetary companion by 16 Cyg A are discussed as
potential explanations. The existence of two other solar-like parent stars,
whose Li (and Be) does not show strong depletion, i.e. whose behaviour is like
16 Cyg A, the Sun and the majority of similar stars with Li and Be abundances
available, implies that the 16 Cyg system requires special observational and
theoretical attention.

A low upper limit has been derived for the beryllium abundance of $\rho ^1$
Cnc. This is the first time a precise limit has been set and that such strong
Be depletion has been observed in a late G-/early K-type MS star. This
measurement clearly constrains the depletion predictions of the available
mixing mechanisms, but requires observation of planet-free stars with similar
age and spectral type to discard the potential effects of the planetary
companion on the Li and Be depletions. Claims have also been made indicating
that $\rho ^1$ Cnc appears to be a subgiant. If this were the case, its strong
Li and Be depletions could be explained by a dilution process taking place
during its post-MS evolution.

\acknowledgements{}
We are indebted to R. Corradi, S. Kemp, B. Bates, and C. R. Benn for obtaining
part of the observations included in this work. We also thank R. Rebolo, J. E.
Beckman, and G. Gonzalez for wide discussions and critical remarks on this
topic. The comments of the referee, Dr. F. Spite, have been of value in
improving the content of this article.

This work has made use of the SIMBAD database and was partially supported by
the Spanish DGES under projects PB92-0434-C02-01 and PB95-1132-C02-01.

\end{document}